\documentclass[prc,preprint,showpacs]{revtex4}
\usepackage{graphicx,dcolumn,array,bm,amsmath,amssymb}

\newcommand{\Imag}{{\cal I}{\rm m}}
\newcommand{\Reg}{{{\cal R}eg}}

\begin{document}

\title{Gamow and R-matrix Approach to Proton Emitting Nuclei}

\author{A.T. Kruppa}

\email{atk@chaos.atomki.hu}

\affiliation{Institute of Nuclear Research, Bem t\'er 18/c, 
4026 Debrecen, Hungary}

\affiliation{Joint Institute for Heavy Ion Research,
Oak Ridge National Laboratory 
\\Oak Ridge, Tennessee 37831}

\author{W. Nazarewicz}

\email{witek@mail.ornl.gov}

\affiliation{Department of Physics and Astronomy, University of Tennessee,
  Knoxville, Tennessee 37996}
  
\affiliation{Physics Division, Oak Ridge National Laboratory, \\ P.O. Box 2008, Oak Ridge,
  Tennessee 37831}
  
\affiliation{Institute of Theoretical Physics, Warsaw University, \\ ul. Ho\.za 69, PL-00681
  Warsaw, Poland} 

\date{\today}

\begin{abstract}
Proton emission from deformed nuclei is described  within  the 
non-adiabatic weak coupling model which takes into account the
coupling to $\gamma$-vibrations around the axially-symmetric  shape. The 
coupled equations are derived
within the Gamow state formalism. 
A new method, based on the combination of the R-matrix theory and the oscillator
expansion technique, is introduced that
allows for a substantial increase of the number
of coupled channels. 
As an example, we study 
the deformed proton emitter $^{141}$Ho.
\end{abstract}

\pacs{23.50.+z, 24.10.Eq, 21.10.Pc, 27.60.+j}

\maketitle

\section{Introduction}

Theoretical models applied to
 the description of non-spherical proton emitters 
 can be divided into two groups. The core-plus-particle models
 describe the radioactive parent nucleus in terms of a single
proton interacting with a core (i.e., the daughter nucleus).
Usually, the core is represented  by 
some phenomenological collective model, e.g., 
 the  Bohr-Mottelson (geometric) 
 model. Depending on the structure of the daughter nucleus,
rotational 
\cite {kru00,bar00,es01} 
or vibrational \cite {dav01,hag01} couplings are assumed. 
The models belonging to this group employ 
the coupled-channel formalism of reaction theory  which has
 been developed in the context of elastic or inelastic scattering.  

Models belonging to the second group employ
the framework of the deformed shell model. In the simplest case,  the 
proton resonance  corresponds to  a Nilsson 
state of a
deformed mean field 
\cite{kad96,fer97,mag98,dav98,mag99,ryk99,son99}. 
Approaches belonging to this group 
can be generalized  
to include the  BCS pairing \cite{fio03}. 

We may refer to  the first group of models as  
weak-coupling models or coupled-channel models.     
For the second group of models,
we reserve the term
resonance Nilsson-orbit (or adiabatic) models.
The term ``adiabatic"   requires an explanation. It is very difficult
 to relate both  groups of models
to each other,  because they operate on different approximation levels. 
In special situations, however,  this relationship
 can be revealed. For instance, in the limit of the infinite moment of inertia
 of the axial  weak-coupling model (which implies degenerate rotational bands and 
  strong rotational coupling  \cite{BMII}), one recovers  
the resonance Nilsson-orbit model \cite{kru03}. 
So one may say that \emph{in this case} 
the adiabatic model is 
an approximation to the weak-coupling (non-adiabatic) picture. 
Generally, however, 
the relation between  adiabatic and non-adiabatic descriptions 
is not  simple. For example,
the resonance  Nilsson-orbit model with  a triaxial  potential
\cite{dav03a} (i.e.,  nonzero $\gamma$ deformation) 
cannot be trivially  related to a weak-coupling model extended
to triaxial degrees of freedom  \cite{kru03}.

If the coupled-channel model with  the rotational coupling is applied to the 
nucleus $^{\rm 141}$Ho, the ground-state decay characteristics 
(half-life time and
branching ratio) are poorly described \cite{bar00,es01}. 
There are several explanations
possible. For example,  it may be that 
the Coriolis mixing is too strong  \cite{es01}. This 
can be partly  cured if pairing is introduced \cite{fio03}. 
Another possibility, explored 
in this work, is the coupling to triaxial vibrations.
Indeed, in particle-plus-rotor 
calculations,  the best description of 
the experimentally observed band structures of  
$^{\rm 141}$Ho can be explained if $\gamma$ deformation is
considered \cite{sew01}. In addition, in the neighboring
nuclei, such as $^{136}$Sm and
 $^{140}$Gd,  there are  low-lying   $2^+_2$  and  
$3^+$ levels \cite{ensdf} which have been
interpreted \cite{wal01}
as members of a $\gamma$-vibrational band.
There are also other indications that in this mass region 
the coupling to triaxial modes can play a role \cite{yan93,kor90}. 
The possibility that triaxiality influences  the decay of 
$^{141}$Ho was investigated in our earlier work \cite{kru03} and 
also in the recent Refs. \cite{dav03a,dav03}
based on  an adiabatic model assuming a 
triaxially deformed mean field.

In this work,  we present   non-adiabatic 
calculations in which  the excitations  of the daughter nucleus 
are properly taken into account. Unlike in Ref.~\cite{dav03},
 we do not assume a 
permanent $\gamma$ deformation of the core, 
 but rather we  consider 
$\gamma$ vibrations around the axially-symmetric deformed shape.

The ground-state  rotational band of $^{140}$Dy  has recently been
observed \cite{kro02,cul02}. In addition, in our work we assume that 
$^{140}$Dy   has the $K$=2  $\gamma$-vibrational band.       
This structure can be coupled to the ground-state band if the proton-daughter
interaction in the body-fixed system deviates from the axial symmetry.
The experimentally observed rotational band of the parent 
nucleus is assumed to be a  $K^\pi$=7/2$^-$ band \cite{sew01}
built upon the [523]$_{\Omega=7/2}$ Nilsson level. In the 
strong-coupling picture the  presence of  the $\gamma$ band in 
$^{140}$Dy implies the existence  of 
two additional   rotational bands in $^{141}$Ho with 
$K$=$\Omega \pm 2$, i.e.,  $K^\pi=3/2^-$ and $K^\pi=11/2^-$.   

In the weak-coupling model, proton emission is described by means
of a 
coupled set of differential equations which are solved assuming
appropriate boundary
conditions.  The most obvious way to describe the proton emission is to 
assume outgoing boundary conditions. This  immediately leads to 
the notion of the Gamow or resonant states, 
the generalized eigenstates of the time-independent Schr\"{o}dinger
equation, which are regular at the origin and satisfy purely outgoing boundary
conditions. Together with non-resonant  scattering states, 
Gamow states form a complete set,
the so-called Berggren ensemble \cite{berggren}, which can be used in
a variety of applications \cite{vertse87}, including the recently
developed Gamow shell model \cite{Mic02,Mic03,betan}.

Unfortunately,  the number of coupled equations rapidly increases with the number
of excited states of the daughter nucleus 
taken into account. In addition, the solution 
of the eigenvalue problem of a very large set of coupled equations 
becomes numerically
unstable at some point. This is especially true if
 one keeps in mind that there is
 a  twenty-order-of-magnitude difference between the real and 
 imaginary part of the energy 
of the Gamow-state which
describes the proton decay of $^{141}$Ho. A possible 
way out is to consider the R-matrix theory. 
However,  even in this case,  one has to deal with
 large sets of coupled differential equations.

In order to avoid the difficulty of solving large sets of 
 coupled differential equations,
one  may use the Rayleigh-Ritz variational principle and apply the basis
expansion method.
In this paper,  the spherical harmonic oscillator wave
functions are used as basis functions.
It was recognized a long time ago that by using the basis
 expansion method the positions 
of narrow resonances can be determined. In particular,
 the signature  of a narrow resonance is that the specific  
positive energy solution is  locally stable with respect to the change 
of the size of the basis 
\cite{Haz70,Rei79,Lov82,Man93,sal96,Kru99,Ara99,kruppa00}.
Several proposals exist in the literature on how to determine
the width of the resonance in this method. They are 
called  $L^2$ stabilization methods \cite{sal96}.
(The name comes from the fact that 
only square integrable functions are used in the expansion.) 
In this paper we will introduce a new method which  
is a combination of  the oscillator expansion method and the R-matrix formalism.
This method is very simple and proves to be accurate enough for  
very narrow proton resonances.
 
The paper is organized as follows. 
We  begin in Sec.~\ref{Wcoupl} with an overview of the weak-coupling model 
applied to the case of
rotational motion and  $\gamma$ vibrations. 
Section~\ref{SecIII} reviews different methods to calculate the position and width 
of a resonance state:
 the theory of Gamow-states,  the 
standard R-matrix formalism, and the new
method which combines the
oscillator expansion method with the R-matrix formalism. 
Finally,   Sec.~\ref{SecIV} contains 
 results of numerical calculations. 
We check the accuracy of the  new method and  
demonstrate  how the position of  
excited states in the daughter nucleus can influence predictions of 
the  weak-coupling model. We also present results 
for the proton emission in  $^{141}$Ho. Finally,
Sec.~\ref{SecCon} contains the conclusions of this work.
 
\section{Weak-coupling model}\label{Wcoupl}
The proton-emitting parent nucleus  is described here in terms
of  a single proton 
coupled to  a deformed core. 
The model Hamiltonian can be written as
\begin{equation}\label{weakh}
H_{\rm rot}=H_{\rm d}-\frac{\hbar^2}{2m}\bigtriangleup_{\bf r}+
V_{\rm def}({\bf r},\omega),
\end{equation}
where $H_{\rm d}$ is the (collective)  Hamiltonian  of the 
daughter nucleus, the second term represents 
the relative proton-daughter kinetic energy,  and $V_{\rm def}$  is the proton-core interaction, which 
depends on the position of the proton ${\bf r}$ and the orientation 
$\omega$ of the core.

\subsection{The proton-daughter interaction}

It is straightforward to define  $V_{\rm def}$
in the body-fixed frame, in which one can define the deformed mean field.
By expanding 
the nuclear radius  in 
multipoles and assuming  quadrupole deformations only, one obtains
\begin{equation}\label{surf}
R(\theta',\phi')=R_0C(a_0,a_2)\left [ 1+a_0Y_{2,0}(\theta')+
a_2( Y_{2,2}(\theta',\phi')+
Y_{2,-2}(\theta',\phi'))\right ]
\end{equation}
where $C(a_0,a_2)$ is the volume conservation factor.
The intrinsic deformed field is defined using a Saxon-Woods form factor 
\begin{equation}\label{sw}
V_{\rm def}(r,\theta'\phi')=-\frac{V_0}{1+\exp\left[(r-R(\theta',\phi'))/a\right]}.
\end{equation}
Expanding to the  first order in 
$a_2$, one obtains
\begin{equation}
V_{\rm def}(r,\theta'\phi')=V_1(r,\theta')+a_2V_2(r,\theta') \left[Y_{2,2}(\theta',\phi')+
Y_{2,-2}(\theta',\phi')\right].
\end{equation} 
The form factor $V_1(r,\theta')$ is the same as (\ref{sw}) except that $a_2$ is
put equal to zero. The form factor of the second term is given by   
\begin{equation}
V_2(r,\theta')=-\frac{V_0 R(\theta',\phi')e^{{r-R(\theta',\phi')}\over a}}
{a\left[1+e^{{r-R(\theta',\phi')}\over a}\right]^2},
\end{equation}
where, again, $a_2$=0 in $R(\theta',\phi')$.
The deformed form factors  $V_1(r,\theta')$ and $V_2(r,\theta')$  still
depend on $a_0$. After perfoming multipole decomposition of 
 $V_1$ and $V_2$, one obtains 
the intrinsic  potential:
\begin{eqnarray}\label{pot}
&V_{\rm def}(r,\theta'\phi')=V^{(1)}_{\rm def}(r,\theta')+
a_2 V^{(2)}_{\rm def}(r,\theta'\phi')\nonumber\\
&=\sum_{\lambda}V_\lambda^{(1)}(r)Y'_{\lambda,0}(\theta')+
a_2 \sum_{\lambda}V_\lambda^{(2)}(r)\left[ Y'_{\lambda,2}(\theta',\phi')+
Y'_{\lambda,-2}(\theta',\phi')\right ].
\end{eqnarray}
For explicit expressions for $V_\lambda^{(1)}(r)$ and $V_\lambda^{(2)}(r)$ 
see, e.g.,  Ref.~\cite{ta65}.
It can be shown that in the laboratory system the daughter-proton 
interaction is given by
\begin{eqnarray}\label{potl}
&V_{\rm def}({\bf r},\omega)=V_{\rm def}^{(1)}({\bf r},\omega)+a_2V_{\rm def}^{(2)}({\bf r},\omega)\nonumber\\
&=\sum_{\lambda \mu}V_\lambda^{(1)}(r)D^\lambda_{\mu 0}Y_{\lambda,\mu}(\hat r)+
a_2\sum_{\lambda\mu}V_\lambda^{(2)}(r)\left( D^\lambda_{\mu 2}+
D^\lambda_{\mu -2}\right )Y_{\lambda,\mu}(\hat r).
\end{eqnarray}
In addition to the nuclear potential, there is also a long-range Coulomb interaction
between the deformed core and the proton. The deformed 
Coulomb form factors, $V_C^{(1)}$ and $V_C^{(2)}$,
are discussed in  Appendix A. 

\subsection{The coupled channel equations}

The  states of the daughter nucleus
are eigenvectors of $H_{\rm d}$. In this work, we adopt
the rotational-vibrational
collective model. The wave functions of the core,  $\phi_{I\mu K}$, are given by the 
standard ansatz \cite{BMII}:
\begin{equation}\label{colwf}
\phi_{I\mu K}=\sqrt{\frac{2I+1}{16\pi^2(\delta_{K,0}+1)}}\left [D^{I*}_{\mu K}+(-1)^I
D^{I*}_{\mu -K}\right ]\chi_{Kn_2}(a_2)\vert {\rm g.s.}\rangle ,
\end{equation}
where $\chi_{Kn_2}(a_2)$ is a $\gamma$-vibrational wave function.
The wave function of the parent nucleus can be written in the weak-coupling  form
\begin{equation}\label{wf}
\Psi^{JM}=\sum_{IK lj} \frac{u^J_{IK lj}(r)}{r}
\Phi_{JMIKlj},
\end{equation}
where the channel function is given by
\begin{equation}\label{chwf}
\Phi_{JMIKlj}=
\sum_{\Omega\mu}\langle j\Omega I\mu\vert JM \rangle {\cal Y}_{lj\Omega}\phi_{I\mu K},
\end{equation}
and
\begin{equation}
{\cal Y}_{lj\Omega}=\sum_{ms}\langle l m\frac{1}{2}s\vert j\Omega\rangle
i^lY_{lm}(\hat r)\chi_{1/2}(s)
\end{equation}
arises from the coupling of the proton spin with the orbital angular momentum.
In our earlier weak-coupling calculations  \cite{kru00,bar00}
there was no summation over $K$ in
Eq.~(\ref{wf}); only 
the $K=0$ term was considered. 
Due to the non-axial symmetric form of the proton-daughter interaction 
(\ref{potl}), the ground state
$K=0$ and the $\gamma$-vibrational $K=2$ band both contribute.

The  radial
functions $u^J_{IK lj}(r)$ are solutions of  the set of coupled-channel equations:
\begin{eqnarray}\label{weakeq}
&\frac{\hbar^2}{2m}\left (-\frac{d^2}{dr^2}+\frac{l(l+1)}{r^2}\right )
u^J_{IKlj}+
\sum_{\lambda I'l'j'} A_\lambda(Ilj,I'l'j',J)B_\lambda(II'K)
V_\lambda^{(1)}u^J_{I'Kl'j'}+\\
&\sum_{\lambda I'K'l'j'} A_\lambda(Ilj,I'l'j',J)C_\lambda(IKI'K',a_2)
V_\lambda^{(2)}u^J_{I'K'l'j'}
=(E-E_{IK})u^J_{IKlj},\nonumber
\end{eqnarray}
where $E_{IK}$ is the energy of the daughter state 
described by the wave 
function (\ref{colwf}).
The $r$-independent coupling coefficients can be written in terms of
the reduced nuclear matrix elements
\begin{equation}\label{redm2}
B_\lambda (II'K)=\langle\phi_{IK}\vert\vert D^\lambda_{;0}\vert\vert\phi_{I'K}
\rangle
\end{equation}
and
\begin{equation}\label{redm3}
C_\lambda (IKI'K',a_2)=\langle\phi_{IK}\vert\vert 
a_2(D^\lambda_{;2}+D^\lambda_{;-2})\vert\vert\phi_{I'K'}
\rangle .
\end{equation}
The explicit expressions for  the geometric coefficients 
$A_\lambda(Ilj,I'l'j',J)$ are given, e.g., in 
Ref.~\cite{tam65}. The nuclear
structure model of the daughter nucleus enters the formalism through the reduced    
matrix elements $B_\lambda$ and $C_\lambda$ \cite{tam65,ta65}.

\section{Calculation of resonance  parameters}\label{SecIII}

The coupled differential equations (\ref{weakeq}) can be 
turned into an eigenvalue 
problem by specifying  boundary conditions. It is always
assumed that the solutions are regular at the origin, i.e.,
$u_c(0)=0$. (From now on, the channel indexes $IKlj$ 
are abbreviated by the symbol $c$.)

\subsection{Gamow states}

To be a Gamow state, the radial wave function must asymptotically
behave as an outgoing Coulomb wave:
\begin{eqnarray}\label{asympwf}
  u_{c}(r) &\stackrel{{\rm large}\; r}{\longrightarrow}
  &O_{l}(\eta,r k_p)\nonumber \\
  &=&G_{l}(\eta,r k_p) + i F_{l}(\eta,r k_p) ,
\end{eqnarray}
where $k^2_c =\frac {2 m}{\hbar^2} ({\cal E}_p-E_{IK})$ and
$\eta k_c=\frac {m} {\hbar^2}Ze^2$.
Such  boundary conditions are only satisfied for a discrete set of 
complex wave numbers $k_c$ which define
the generalized eigenvalues $E={\cal E}_p$ of  Eq.~(\ref{weakeq}).
These  eigenvalues correspond to the poles of the scattering
matrix~\cite{[Hum61],vertse87}.
The corresponding solutions are either bound 
states with negative real energies ${\cal E}_p=E_b<0$ and pure 
imaginary wave numbers $k_p=i\gamma_p$ ($\gamma_p>0$), 
or resonance states,
${\cal E}_p=E_{\rm res} -i \frac{\Gamma_{\rm res}} {2}$, with  nonzero 
imaginary parts
$\Gamma_{\rm res} \neq 0$, and $k_p=\kappa_p-i\gamma_p$.
 
The asymptotic behavior of the radial wave functions are determined by $k_p$.
For Gamow states these functions show
oscillating behavior at large values of $r$ so one  must define a new
normalization scheme.  
Berggren proposed \cite{berggren} a generalized scalar product 
and introduced a regularization
procedure ($\Reg$).  With this generalization the norm is
\begin{equation}
\label{sphnorm}
\sum_{c} \Reg \int_0^{\infty} [u_{c}(r)]^2 \, dr 
= 1.
\end{equation}

Once   the resonance energy and radial wave function have been determined,
 there are 
different methods to calculate the width of the state.  The simplest
 method is to take twice the imaginary part of the energy of the 
resonance.  However, for narrow resonances
the accurate numerical calculation of $\Imag [{\cal E}_p]$ 
is difficult.
Therefore, other  methods are often used. One possibility is
to  calculate the partial width for each channel from the so-called current
expression~\cite{[Hum61]}
\begin{equation}
\label{partcur}
\Gamma_{c}(r)=i\frac{\hbar^2}{2\mu}  
\frac{u_{c}^{\prime*}(r) u_{c}(r) 
- u^{\prime}_{c}(r) u_{c}^*(r)}
{\sum_{c^{\prime}} \int_0^{r} |u_{c^{\prime}}(r')|^2 
  {\rm d}r'} ,
\end{equation}
where the sum of the partial widths
\begin{equation}
\label{totgam}
\Gamma_{\rm res} = \sum_{c} \Gamma_{c}(r) 
\end{equation}
gives the total  decay width. Although values of $\Gamma_{c}(r)$ depend
on $r$ in the region where the coupling potential terms are not
negligible,  the total width (\ref{totgam}) is  independent of $r$, 
which reflects flux conservation. 

In practice, the Gamow boundary condition given by Eq.~(\ref{asympwf}) 
can be implemented  in the form 
\begin{equation}\label{gbcond}
\frac {u'_c (r_{\rm as})}{u_c (r_{\rm as})}=k_p\frac{ 
O'_{l}(\eta ,r_{\rm as} k_p)}{O_{l}(\eta ,r_{\rm as} k_p)},  
\end{equation}
where $r_{\rm as}$ is the channel radius (the off-diagonal couplings 
are negligible for $r > r_{as}$). Using   Eq.~(\ref{gbcond}),
the partial decay width can be
written at the point $r_{as}$ as
\begin{eqnarray}
\label{gwidth}
\Gamma_{c}(r_{\rm as})& = &i\frac{\hbar^2 }{2\mu}  
\frac{|u_{c} (r_{\rm as})|^2}
{|O_{l}(\eta,k_p r_{\rm as})|^2 
\sum_{c'} \int_0^{r_{\rm as}} |u_{c'}(r')|^2 
  {\rm d}r'}\nonumber\\
  & \times & \left[ k_p^{*} O'^*_{l}(\eta,r_{\rm as}k_p) 
  O_{l}(\eta,r_{\rm as}k_p) 
 -  k_p O'_{l}(\eta  ,r_{\rm as}k_p) 
  O^*_{l}(\eta ,r_{\rm as} k_p) \right].
\end{eqnarray}
If one neglects the imaginary part of $k_p$, the square bracket 
in Eq.~(\ref{gwidth}) becomes
$-2i$ and the  expression for the partial decay width
can be written in a simple form:
\begin{equation}\label{gwidth2} 
\Gamma_{c}(r_{\rm as}) \approx \frac{\hbar^2 \kappa_p}{\mu}  \frac{ 
|u_{c} (r_{\rm as})|^2}
{|O_{l}(\eta, k_p r_{\rm as})|^2
\sum_{c'} \int_0^{r_{\rm as}} |u_{c'}(r')|^2
  {\rm d}r'}.
\end{equation}
Equation
(\ref{gwidth}) and its approximate form (\ref{gwidth2}) are strictly 
valid only
at the point $r_{\rm as}$ where the boundary condition is given.
We emphasize at this point that if the coupled
equations are solved with the Gamow boundary condition, then the total
width can be calculated at any value of $r$  using {\em exact} relations  
(\ref{partcur}) and (\ref{totgam}).

\subsection{R-matrix method}

For completeness,  we summarize those important aspects of
the R-matrix theory \cite{La58} which 
are  relevant to our work.
In the R-matrix theory one also deals with  a set of radial functions $g_c(r)$. 
These functions are 
regular at the origin and satisfy the same coupled equations (\ref{weakeq}) 
as the Gamow states but with the following boundary conditions 
 \begin{equation}\label{rbcond}
a\frac{g_c'(a)}{g_c(a)}=B_c,
\end{equation}  
where the parameters $B_c$ are arbitrary 
real numbers. It is assumed that
the short-range diagonal and off-diagonal proton-core interactions 
can be neglected beyond the channel radius $a$. Consequently,
 $a$ has the same meaning as the parameter $r_{\rm as}$  
of the Gamow theory. It is worth noting, however, that
$a$ is always real, while $r_{\rm as}$ can be complex.
 
The boundary 
condition (\ref{rbcond}) defines the complete set of functions 
inside the channel surface. The 
real eigenvalues of the coupled-channel equations  
are denoted by $E_\lambda$ and the corresponding 
eigenfunctions by $g^\lambda_c(r)$. They are normalized to one 
inside the channel surface, 
\begin{equation}\label{rnorm} 
\sum _c \int_0^a |g^\lambda_c(r)|^2 dr=1,
\end{equation} 
and define the so-called reduced width amplitudes
\begin{equation}\label{rwidth}
\gamma_{\lambda c}=\left(\frac{\hbar ^2}{ 2m_c a}\right)^{1/2}g^\lambda_c(a).
\end{equation}
The resulting R-matrix has a  simple form 
\begin{equation}\label{rmat}
R_{cc'}(E)=\sum_\lambda \frac{\gamma_{\lambda c}\gamma_{\lambda c'}} 
{E_\lambda-E}
\end{equation}
but it is related to the physically important scattering S-matrix in 
a complicated way  \cite{La58}. Let us emphasize that
the  calculated S-matrix
is independent from both the boundary condition parameters $B_c$ and from 
the channel radius $a$  {\em only} if all the R-matrix states
are taken into account in Eq. (\ref{rmat}).

Assuming that in a given energy region only one term 
dominates in the R-matrix and making further approximations 
(see p. 322 of Ref. \cite{La58}), Lane  and Thomas showed 
that the S-matrix can be written in the form
\begin{equation}\label{smat}
S_{cc'}(E)\approx S^0_{cc'}(E)\left [\delta_{c,c'}+
\frac{i\Gamma_{\lambda c}(E)^{1/2}\Gamma_{\lambda c'}(E)^{1/2}}
{E_\lambda+\Delta_\lambda(E)-E-\frac{i}{2}\Gamma_\lambda (E)}\right ],
\end{equation}
where
the partial R-matrix widths
\begin{equation}
\Gamma_{\lambda c}(E)=2 P_{l_c}(E)\gamma_{\lambda c}^2
\end{equation}
give the total width 
\begin{equation}
\Gamma_\lambda(E)=\sum_c \Gamma_{\lambda c}(E).
\end{equation}
In Eq.~(\ref{smat}),  function $\Delta_\lambda (E)$ is given by 
\begin{equation}
\Delta_\lambda(E)=\sum_c\Delta_{\lambda c}(E),
\end{equation}
where 
\begin{equation}
\Delta_{\lambda c}(E)=-\left (S_{l_c}(E)-B_c\right )\gamma^2_{\lambda c}.
\end{equation}
The penetration $P_{l_c}(E)$ and shift $S_{l_c}(E)$ functions 
are related to the Coulomb $F_{l_c}$ and $G_{l_c}$ functions (see p. 270 of Ref. 
\cite{La58}).

Within  approximation (\ref{smat}), the complex-energy resonance poles of the S-matrix,
${\cal E}^R_p=E^R_{\rm res}-\frac{i}{2}\Gamma^R_{\rm res}$, 
satisfy 
 the equation 
\begin{equation}\label{req}
E_\lambda+\Delta_\lambda({\cal E}^R_p)-{\cal E}^R_p-
\frac{i}{2}\Gamma_\lambda ({\cal E}^R_p)=0.
\end{equation}
Here, we used the upper index $R$ in order to distinguish 
this R-matrix approximation for the resonance energy from the energy of the 
corresponding Gamow state. In order to simplify the solution of the non-linear 
equation (\ref{req}), one often introduces further  approximations 
and assumptions for the calculation of the functions $\Delta_\lambda (E)$ and
$\Gamma_\lambda (E)$.

In the method of Thomas \cite{Th54},  the function 
$\Delta_{\lambda c}(E)$ is expanded 
around the R-matrix eigenvalue $E_\lambda$ 
\begin{equation}\label{expansion}
\Delta_{\lambda c}(E)\approx -(S_{l_c}(E)-B_c)-\dot S_{l_c}( E_\lambda)
(E-E_\lambda ), 
\end{equation}
where the dot denotes energy derivative. Furthermore,
the $E$-dependence of $\Gamma_{\lambda c}(E)$ is neg\-lected 
and $\Gamma_{\lambda c}(E)$ is replaced
by the corresponding value at $E_\lambda$. Under these
assumptions one obtains
\begin{equation}\label{e1}
E^R_{\rm res}=\frac{E_\lambda+
\sum_c\left (B_c-S_{l_c}(E_\lambda)\right)\gamma_{\lambda c}^2}
{1+\sum_c\dot S_{l_c}(E_\lambda)\gamma_{\lambda c}^2}
\end{equation} 
and
\begin{equation}\label{g1}
\Gamma_{\rm res}^R=\frac {\Gamma_\lambda(E_\lambda)}
{1+\sum_c\dot S_{l_c}(E_\lambda)\gamma_{\lambda c}^2}.
\end{equation}
In order to simplify (\ref{e1}) we may require that the 
chosen boundary condition parameters satisfy the condition:
\begin{equation}\label{ltb}
B_c=S_{l_c}(E_\lambda).
\end{equation}
 If the $\dot S_{l_c}(E_\lambda)$ terms are negligible, then 
 the resonance 
energy  corresponds to the R-matrix eigenvalue 
\begin{equation}
E_{\rm res}^R=E_\lambda\label{e2}
\end{equation}
and the width can be 
calculated with the well-known  expression 
\begin{equation}\label{g2}
\Gamma_{\rm res}^R=\sum_c 2 P_{l_c}(E_\lambda)\gamma_{\lambda c}^2.
\end{equation}

Two variants of Thomas's procedure  
can be found in a later paper of Lane and Thomas 
\cite{La58} where they give different expressions for $E_{\rm res}^R$ and 
$\Gamma_{\rm res}^R$.

\subsection{R-matrix method using oscillator expansion}

In this section we propose a simple method, based on the R-matrix formalism,
 to estimate 
the parameters of a resonance. The advantage of this method is that 
it  avoids 
solving   a large set of coupled differential equations. 
The method is based on the expansion of 
  the radial functions $u_{IKlj}(r)$  in the single-particle
  basis  $\phi^{HO}_{nl}(r)$ of the spherical  harmonic oscillator. 
In this basis, 
the total wave function (\ref{wf}) can be written in the form:
\begin{equation}\label{wfho}
\Psi^{JM}=\sum_{IK lj} \sum_n C^J_{IKnlj} \frac {\phi^{HO}_{nl}(r)} {r}
\Phi_{JMIKlj}.
\end{equation}
The coefficients $C^J_{IKnlj}$ can be obtained from the 
 matrix eigenvalue
equation:
\begin{eqnarray}\label{weakeqho}
&\sum _{n'}
\langle\phi^{HO}_{nl}\vert
\frac{\hbar^2}{2m}\left (-\frac{d^2}{dr^2}+\frac{l(l+1)}{r^2}\right )
\vert\phi^{HO}_{n'l}\rangle C^J_{IKn'lj}-(E^{HO}_\lambda-E_{IK})C^J_{IKnlj}\nonumber\\
&+\sum_{\lambda I'n'l'j'}
A_\lambda(Ilj,I'l'j',J)B_\lambda(II'K)
\langle\phi^{HO}_{nl}\vert V_\lambda^{(1)}\vert\phi^{HO}_{n'l'}
\rangle C^J_{I'Kn'l'j'}\\
&+\sum_{\lambda I'K'n'l'j'}A_\lambda(Ilj,I'l'j',J)C_\lambda(IKI'K',a_2)
\langle\phi^{HO}_{nl}\vert V_\lambda^{(2)}\vert\phi^{HO}_{n'l'}\rangle
C^J_{I'K'n'l'j'}=0.\nonumber
\end{eqnarray}
In the following,  the corresponding real eigenvalues 
are denoted as  $E^{HO}_\lambda$.

In the R-matrix theory,  the 
coupled equations (\ref{weakeq})  are solved with  imposed
boundary conditions (\ref{rbcond}).  However, as discussed in the following,
this procedure can be reversed. In the first step, 
we solve the algebraic
eigenvalue problem (\ref{weakeqho}) 
for the coefficients $C^J_{IKnlj}$.
The resulting  radial functions $g_c(r)$ 
define the  {\em boundary condition function}  at the point $r$:
\begin{equation}\label{bfun}
B_c(r)=B_{IKlj}(r)=r \left (\sum_n C^J_{IKnlj} \phi^{HO}_{nl}(r)\right )'/ 
\sum_n C^J_{IKnlj} \phi^{HO}_{nl}(r). 
\end{equation}
Having  determined the boundary 
condition parameter at each $r$,  the 
R-matrix formalism can now be applied.
In particular, after  replacing
 $E_\lambda$ with $E^{HO}_\lambda$ in 
  expressions (\ref{e1}) and (\ref{g1}),
they  can be used
to  compute the position and the width of a resonance at each value of $r$: 
\begin{equation}
E^{HO}_{\rm res}(r)=\frac{E^{HO}_\lambda+
\sum_c\left (B_c(r)-S_{l_c}(E_\lambda^{HO})\right)\gamma_{\lambda c}(r)^2}
{1+\sum_c\dot S_{l_c}(E_\lambda^{HO})\gamma_{\lambda c}(r)^2}
\end{equation} 
and
\begin{equation}\label{GRMHO}
\Gamma_{\rm res}^{HO}(r)=\frac {\Gamma_\lambda(E_\lambda^{HO})}
{1+\sum_c\dot S_{l_c}(E_\lambda^{HO})\gamma_{\lambda c}(r)^2},
\end{equation}
where the $r$-dependent reduced width amplitudes (\ref{rwidth}) are
given by
\begin{equation}
\gamma_{\lambda c}(r)=\left(\frac{\hbar ^2}{ 2m_c r}\right)^{1/2}
\sum_n C^J_{IKnlj} \phi^{HO}_{nl}(r).
\end{equation}
This algorithm is further referred to as the {\em R-matrix method based on 
harmonic oscillator expansion} (RMHO). In RMHO,
the energy  
and  width of the resonance  explicitly 
depend on $r$. However, for  sufficiently large values of $r$,
 this dependence is expected to be
extremely weak.
It is to be noted that since   expression 
 (\ref{g2}) is derived 
under specific assumption (\ref{ltb}), it 
is not valid  in the RMHO method.

The derived boundary condition parameters (\ref{bfun})   
do not depend on the actual normalization used. However,  this is
no longer  true for the reduced width 
amplitudes (\ref{rwidth}).
In order to 
apply the R-matrix method 
at each $a$=$r$, the radial functions 
\begin{equation}\label{radialg}
g_c(r)=g_{IKlj}(r)=\sum_{n=0}^{n_{\rm max}} C^J_{IKnlj} \phi^{HO}_{nl}(r)
\end{equation}
have to be renormalized to one 
inside the channel surface according to 
Eq. (\ref{rnorm}).

\section{Results}\label{SecIV}

The numerical tests have been carried out for the deformed
proton emitter 
$^{141}$Ho, viewed  as a proton-plus-core system, with  
the daughter nucleus $^{140}$Dy  being the collective core.
We employed the same successful   parameterization of the
Woods-Saxon (WS) optical potential as  in earlier Ref.~\cite{bar00}.

\subsection{Resonance Width in RMHO}

Let us first assume that  the core is axially deformed
($a_2=0$).  In the calculations, all the  states
in  the  g.s. rotational band in the daughter nucleus up to 
 $I$=12 were considered.
 In our weak-coupling calculations,
the experimental excitation energies of $^{140}$Dy were used 
for states with $I$$<$10,  and 
the energies of the remaining  states 
were obtained by the variable-moment-of-inertia (VMI)  fit to the  data.
That is, for the g.s. band we took the values:
0.203, 0.567, 1.044, 1.597, 2.218,  and 2.894 MeV.
The deformation parameter $a_0$ was set to the value of
0.244,  which is consistent with earlier investigations 
\cite{sew01,kro02}. The WS  strength was adjusted to reproduce   
the experimental  position of the $J^\pi$=7/2$^-$
resonance at 1.19 MeV.
The number of coupled
channels in this variant  is   46. 
This number  is sufficiently small to 
carry out the reliable calculation of the Gamow-state energy eigenvalue. 
The resulting resonance width is 
 ${\rm 0.208\times 10^{-19}}$ MeV. 
We accept this number as the exact, or reference,  value.

\begin{figure}
\begin{center}
\includegraphics[scale=0.7]{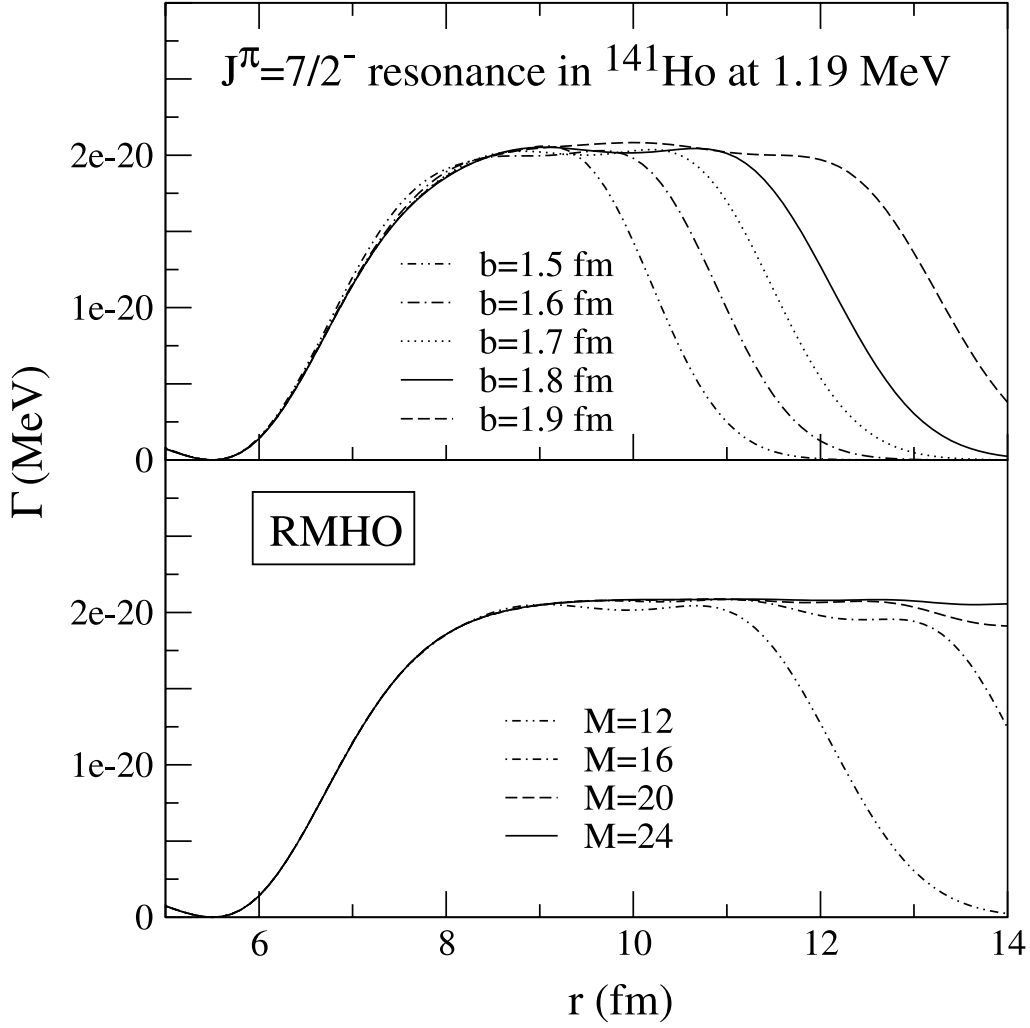}
\end{center}
\caption{\label{figrmho} The width of the $J^\pi$=7/2$^-$
resonance in $^{141}$Ho at  1.19 MeV calculated in RMHO
as a  function of $r$. Top: dependence on the oscillator length
parameter $b$ (the number of basis states is $M$=12). Bottom:
dependence on $M$ ($b$=1.8 fm).
}
\end{figure}
The harmonic oscillator basis is characterized by a single parameter,
the oscillator length $b$.
The upper part of  Fig. \ref{figrmho} shows the resonance width 
(\ref{GRMHO}) calculated in
RMHO as a function of $r$. 
For  each partial wave, $M$=$n_{\rm max}$+1=12 harmonic oscillator  
functions were used in the expansion (\ref{radialg}) and  
the value of $b$  was varied. As  expected, a clear plateau appears at 
large values of   $r$.
The extent of the plateau depends on the size of $b$:
 the greater oscillator  length (i.e., the r.m.s. oscillator radius),
  the greater  the extent 
of the plateau.
The reason for the rapid decrease  of the 
width function $\Gamma^{HO}_{\rm res}(r)$  
at very  large  values of $r$ 
lies in   the fact that the radial channel function is approximated 
by a linear combination of a finite number of oscillator functions,
each having the Gaussian asymptotic behavior. Therefore, by
increasing the number of states in the basis, 
the extent of the plateau is expected to increase. 
This is illustrated in  Fig.
\ref{figrmho} (lower portion) which shows
RMHO results obtained at a fixed value of $b$=1.8\,fm for several
values of $M$. It is seen that for $M$=24
($n_{\rm max}$=23) the width function becomes
 independent of  $r$ in a very wide interval of $r$.
In the interval between $r$=9 and 12~fm the RMHO width exhibits 
tiny  oscillations (practically invisible in Fig. \ref{figrmho}).
Therefore, to obtain a well-defined value, 
 we divide this interval 
equidistantly with a step size  of 
0.1 fm and calculate the average  $\bar{\Gamma}^{HO}_{\rm res}=
\frac{1}{N_r}\sum _{i=1}^{N_r}\Gamma^{HO}_{\rm res}(r_i)$,
which will be considered as the RMHO width in the following.

\begin{figure}
\begin{center}
\includegraphics[scale=0.5]{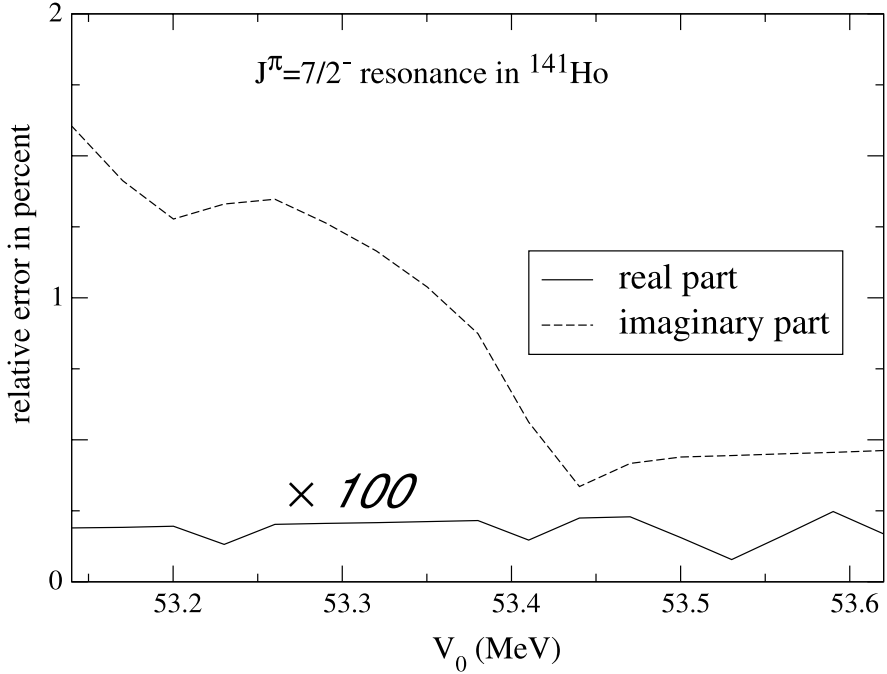}
\end{center}
\caption{\label{error} 
The relative error of the real (solid line) and imaginary 
(dashed line) 
energy of  the $J^\pi$=7/2$^-$
resonance in $^{141}$Ho  calculated in RMHO ($M$=24)
as a  function of the WS potential depth $V_0$.
The reference (exact) values are taken from the Gamow states calculation.
In the considered range of $V_0$,
the resonance width 
changes by four orders of magnitude. Note that the solid curve
has been multiplied by a factor of  100.}
\end{figure}
In order to assess the quality of the RMHO method, 
Fig. \ref{error} shows  the 
relative errors of the real and  imaginary part of the
energy of the  resonance 
as a function of the WS potential depth $V_0$. 
The reference values  were obtained by the  Gamow-state coupled-channel  procedure.
In the considered  region of $V_0$, the resonance width 
changes by four orders of magnitude;  however,  the relative error of RMHO 
is less than 1.7 percent. The accuracy of 
RMHO for the real part of the energy is much better: 
the relative error is always smaller than 0.0025 percent.  
The results presented
in  Figs. \ref{figrmho} and \ref{error} convincingly demonstrate that the RMHO 
formalism can be safely used to calculate isolated  narrow proton resonances. 
 
\subsection{Proton decay of $^{141}$Ho}

In this section we investigate the influence of $\gamma$ vibrations
on  the process of proton emission from $^{141}$Ho.
All results presented in this 
section are obtained with the RMHO method using $M$=20 oscillator 
functions for each partial wave. The oscillator length was assumed to be  $b$=1.8 fm.
Using the results of the VMI fit for the g.s. band, 
the assumed energies of the members of the $\gamma$ band are:
0.750, 0.934, 1.144, 1,378, 1.633, 1.907, 2.198, 2.504, 2.825
3.159, and 3.507 MeV  for $I=2,3,4,\ldots 12$. The chosen 
position of  the $2^+_2$ band head  of the $\gamma$-vibrational band, 750 keV,
was taken according to the systematic trends around   $N$=74.  
 
When  one includes 
the $K$=2 $\gamma$-vibrational band in addition to the g.s. band,
the number of coupled channels increases from 46 to 130 (assuming that the
maximum spin  is $I_{\rm max}$=12 in both bands). For that reason,
we decided to carry out the RMHO calculations instead of the weak-coupling Gamow
analysis.  We have checked, however, that for $I_{\rm max}$=10, where the
coupled-channel calculations with the $K$=2 band can be done, the RMHO results
coincide with those of the  coupled-channel method.

\subsubsection{Structure of $J^{\pi}$=7/2$^-$ states}

The simplest model of the g.s. decay of $^{141}$Ho is   based on the adiabatic resonant 
Nilsson-orbit picture of Sec. II.B of Ref.~\cite{bar00}. Here, the valence proton 
occupies a $\Omega^\pi$=7/2$^-$ Gamow state in an axially deformed mean field. 
Let us  consider this scenario first.
 In our WS model,  there is only
one $\Omega$=7/2$^-$ state in the energy region $\rm (-3\ MeV,1\ MeV)$
and  deformation ($a_0$$\sim$0.244, $a_2$=0).
The calculated energy of this [523]7/2 state is 0.426 MeV. 
There are
three more negative parity Nilsson states originating from the $h_{11/2}$ proton 
intruder shell,
with energies  --2.678 MeV ([550]1/2), --2.141 MeV ([541]3/2),  and --1.105
MeV ([532]5/2). 
If we now apply  the weak-coupling model 
(i.e., we assume that the daughter nucleus
 has a g.s. rotational  band with the finite moment of inertia),
we calculate one 
 $J^{\pi}$=1/2$^-$ state,
two 3/2$^-$ states,  three 5/2$^-$ states, 
and four 7/2 $^-$ states in the considered energy region.
Of  those four 7/2$^-$ states,
 only one 
can be associated with the  7/2$^-$ band-head from which the proton emission takes place.
The remaining three are rotational excitations
associated with the
$K_{\rm par.}$=$\Omega$=1/2, 3/2, and 5/2 bands  built upon the deformed Nilsson levels
mentioned above. 
Table \ref{omega1} displays the structure of the  $J^{\pi}$=7/2$^-$ states
calculated in the non-adiabatic approach.
The $\Omega$ decomposition of the states 
\cite{bar00}  clearly identifies  
the Nilsson orbit upon  which the rotational g.s. band of the parent nucleus is built.

\begin{table}
\caption{\label{omega1}$\Omega$ decomposition of the $J^{\pi}$=7/2$^-$ 
states in $^{141}$Ho in the energy region (-3 MeV, 1 MeV) 
calculated in the non-adiabatic approach.  The axial deformation 
($a_0$=0.244, $a_2$=0) is assumed.} 
\begin{tabular}{|c|cccc|}
\hline
$E_r$ (MeV)  & $\Omega$=1/2 & $\Omega$=3/2  &$\Omega$=5/2 &
 $\Omega$=7/2 \\
\hline
-2.255 & 0.828 & 0.163 & 0.009 & 0.000\\
-1.066 & 0.168 & 0.694 & 0.135 & 0.003\\
-0.103 & 0.006 & 0.144 & 0.808 & 0.042\\
 1.190 & 0.000 & 0.001 & 0.045 & 0.954\\
 \hline
\end{tabular}
\end{table}

The situation becomes more complex if, in addition to the g.s. band,
 one also considers
 the $K$=2 rotational band in
the daughter nucleus (i.e., if one takes the non-zero triaxial coupling $a_2$).
The coupling to $\gamma$ vibrations immediately results in an increase
of the numbers of predicted bands. Indeed, since the $\gamma$ band
can be built upon each $K_{\rm par.}$=$\Omega$ structure, one obtains twelve
bands  with quantum numbers $K_{\rm par.},\ K_{\rm par.}+2$, and   
$\vert K_{\rm par.}-2\vert $ in the  energy interval considered.
Among those twelve bands, only two  have a 
$J^{\pi}$=7/2$^-$ band head. One
is the previously discussed [523]7/2 band while the other corresponds to
a $\gamma$-phonon built upon the [541]3/2 Nilsson orbital.
Table \ref{omega2} displays the  $\Omega$ decomposition of the four 
states of Table \ref{omega1}  
 in the presence of a small $\gamma$ coupling ($a_2$=0.05). It is seen
 that the single-proton band head is clearly identified.
\begin{table}
\caption{\label{omega2}Same as in Table~\protect\ref{omega1} but in the presence of 
small triaxial  coupling ($a_2=0.05$).} 
\begin{tabular}{|c|c|c|c|c|c|c|}
\hline
$E_r$&\multicolumn{4}{c|}{$K$=0}&\multicolumn{2}{c|}
{$K$=2}\\ \cline{2-7}
(MeV) & $\Omega$=1/2 &
$\Omega$=3/2 & $\Omega$=5/2  & $\Omega$=7/2  & $\Omega$=1/2  &
$\Omega$=3/2 \\
\hline
-2.616&0.691&0.232&0.029&0.007&0.034&0.007\\
-1.184&0.015&0.323&0.004&0.014&0.429&0.215\\
0.122&0.028&0.078&0.529&0.214&0.134&0.017\\
1.153&0.015&0.003&0.018&0.902&0.002&0.060 \\
\hline
\end{tabular}
\end{table}

\subsubsection{Proton emission from the ground state of $^{141}$Ho}

\begin{figure}
\begin{center}
\includegraphics[scale=0.5]{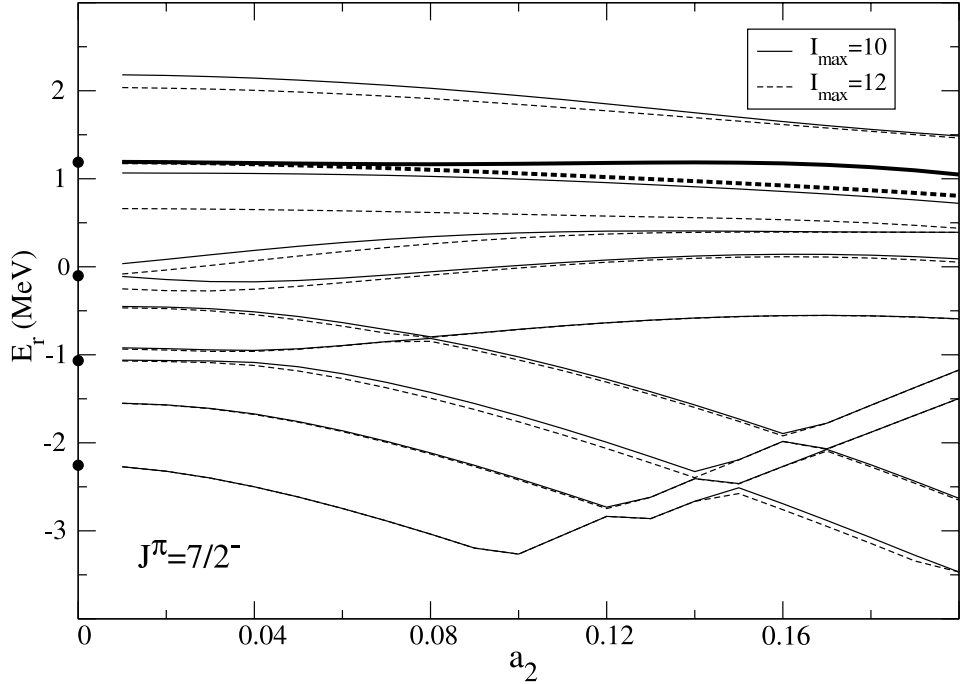}
\end{center}
\caption{\label{conv} The position of the 
 bound and resonance 
$J^{\pi}$=7/2$^-$ states in $^{141}$Ho  calculated in the weak coupling
model as  a
function of the triaxial coupling constant  $a_2$ for 
$I_{\rm max}$=10 and 12  ($I_{\rm max}$ is the maximum value of angular momentum 
considered in the  g.s. band and in the gamma  band
of the daughter nucleus $^{140}$Dy). The results for $I_{\rm max}$=12 
are fully converged, i.e., a further increase of the number of states
does not change results in the scale of this figure.
The axial results without coupling the $\gamma$ band
(cf. Table.~\protect\ref{omega1})  are marked by the dots.
The 7/2$^-$ ground-state  of $^{141}$Ho is marked by thick lines.}
\end{figure}

Earlier investigations \cite{bar00,kru03} have demonstrated that
in the weak coupling model  there is a sensitivity of
 the resonance's parameters  to the  number of states 
 in the rotational bands of the daughter nucleus
 taken into
account. Figure \ref{conv}
shows calculated energies of the  $J^{\pi}$=7/2$^-$ states  in $^{141}$Ho
as a function of the coupling constant $a_2$ ($a_0=0.244$) 
for several values of $I_{\rm max}$. 

Figure \ref{conv}
shows that for bound states the convergence is already very satisfactory for   
$I_{\rm max}$=10; however,  this is not true for the 7/2$^-$ band head.
For instance, at  $a_2$=0.1
 one obtains for the energy of lowest  bound state:
-3.254, -3.264, -3.265, -3.265, and -3.265 MeV
for $I_{\rm max}$=8, 10, 12, 14,
and 16, respectively. 
In contrast,  for the  7/2$^-$ g.s. (marked by thicker lines in Fig.   \ref{conv})
the analogous numbers are: 
1.297, 1.171, 1.062, 1.062, and 1.062 MeV. That is, 
 in this case,  going from 
$I_{\rm max}$=10 to $I_{\rm max}$=12 the energy changes by as much
as 109 keV. This variation is significant since  the width of the 
resonance is 
extremely  sensitive to its  energy.
In our previous  paper  \cite{kru03},  we made the  pilot studies of the
coupling to   the $\gamma$ band on proton emission 
in $^{141}$Ho. Unfortunately, in this
early  analysis based on the coupled-channel method,  we 
took  $I_{\rm max}$=10; hence the  conclusions of this paper have to be revised
 (see below).

\begin{figure}
\begin{center}
\includegraphics[scale=0.5]{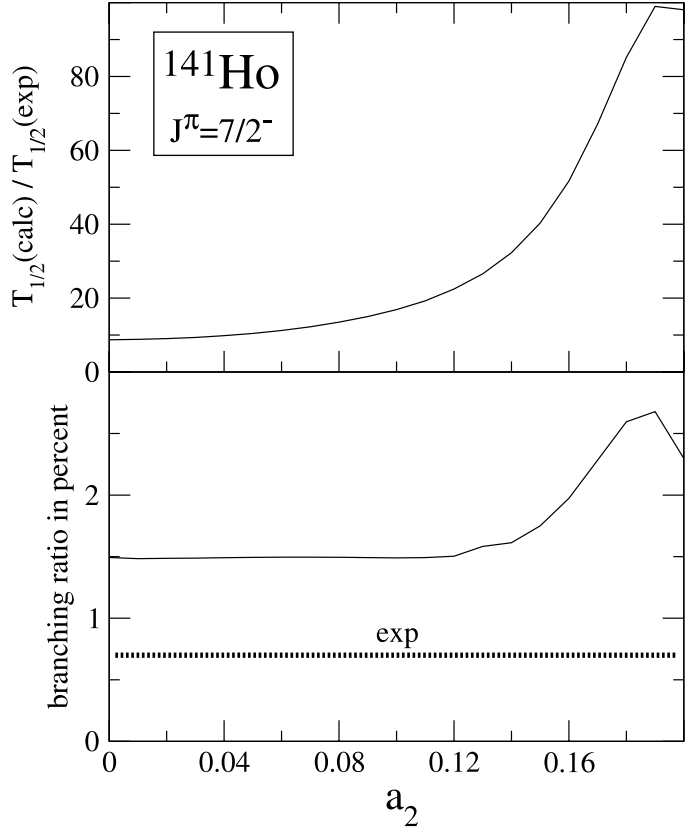}
\end{center}
\caption{\label{width}Half-life (top)  and branching ratio for the decay 
to the $2^+_1$ state
in $^{140}$Dy (bottom) 
for the 7/2$^-$  g.s. of  $^{141}$Ho as a 
function of the triaxial coupling  $a_2$. Experimental data
($T_{1/2}$=4 ms and branching ratio 0.7\%)  are taken
from Refs. \protect\cite{ryk99,Data141Ho}. The spectroscopic factor (BCS 
occupation coefficient)  was  assumed to be $u^2$=0.84 \protect\cite{bar00}.}
\end{figure}

The width of the   7/2$^-$ band head  was computed  using the RMHO method
assuming  $I_{\rm max}$=12. At each value 
of $a_2$ we have adjusted the potential depth 
so as to get the position of the resonance at 1.19 MeV. The 
calculated half-life of the 
resonance and the branching ratio for the decay to the $2^+_1$ state
in $^{140}$Dy are displayed in Figure \ref{width}. It is seen that when increasing
 the coupling to the $\gamma$-band, both the lifetime and 
 the $2^+_1$ branching ratio increase, and the agreement with experiment gets worse.

\begin{figure}
\begin{center}
\includegraphics[scale=0.5]{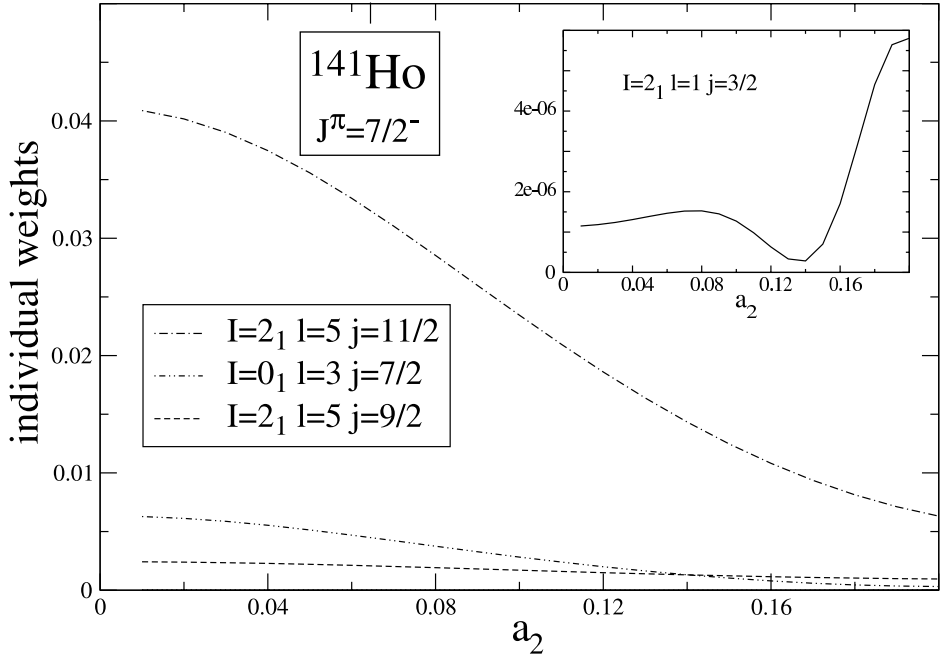}
\end{center}
\caption{\label{weight}Weights (\protect\ref{Clj}) 
of the ($I$=2$_1$, $l$=5 $j$=11/2),  (0$_1$, 3, 7/2), (2$_1$, 5, 9/2),
  and (2$_1$, 1, 3/2)  
partial waves of  the 7/2$^-$  g.s. of $^{141}$Ho
 as a function of $a_2$.}
\end{figure}

In order to understand the behavior shown in Fig.~\ref{width},
we analyzed the components of 
the wave function. Figure \ref{weight} shows  the   
weights of various partial waves ($IKlj$)
in the  7/2$^-$  g.s. of  $^{141}$Ho,
\begin{equation}\label{Clj}
|C_{IKlj}|^2=\int_0^\infty u^2_{IKlj}(r) dr,
\end{equation}
as  functions of  $a_2$. According to our calculations, the amplitudes
associated with the coupling to the
$0^+_1$ g.s. and $2^+_1$ state in $^{140}$Dy are fairly small; most of
the strength lies in higher-lying states including the  channels
that are energetically closed for proton emission. The ($0^+_1$, $f_{7/2}$)
 amplitude,
solely determining the 7/2$^-$$\rightarrow$$0^+_1$ decay, 
gradually decreases with $a_2$. Interestingly, while the {\it total} $f_{7/2}$
strength {\it increases} with $a_2$ as expected (the $h_{11/2}$ and 
$f_{7/2}$ shells are strongly coupled by triaxial field), most
of this strength is pushed up to higher-lying states. 

\begin{figure} 
\begin{center}
\includegraphics[scale=0.5]{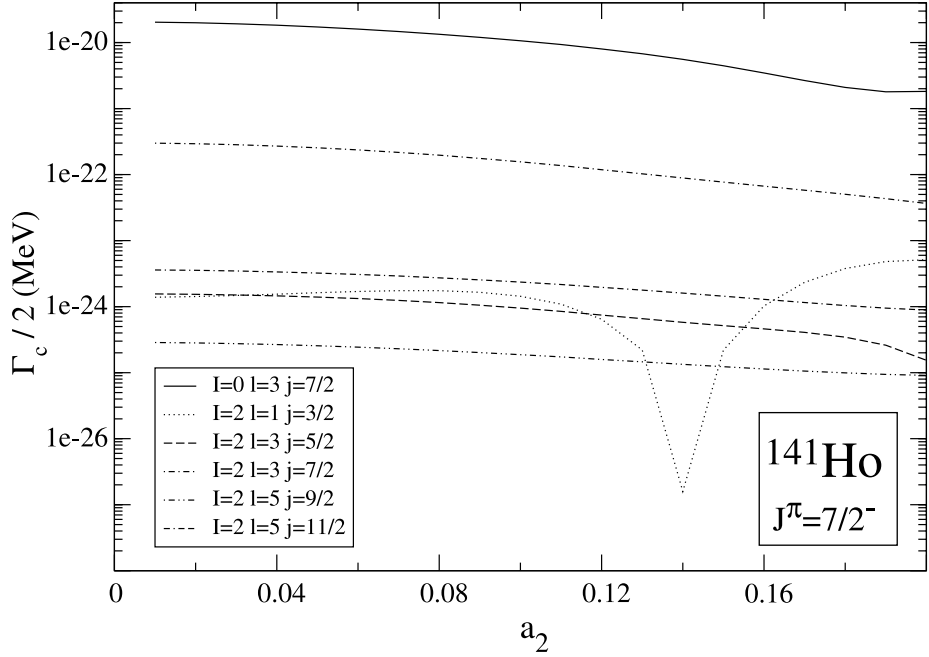}
\end{center}
\caption{\label{gammacc}Partial widths $\Gamma_c$
(\ref{partcur}) corresponding to various  
$0^+_1$ and $2^+_1$ decay channels 
 as  functions of $a_2$.}
\end{figure}

Figure \ref{gammacc} displays  partial widths $\Gamma_c$
(\ref{partcur}) corresponding to various  channels of  decay
to the $0^+_1$ and $2^+_1$ states in $^{140}$Dy. The gradual decrease
of  the (0$_1$, $f_{7/2}$) partial width (hence the increase of
the  half-life of $^{141}$Ho) with $a_2$ can be explained
in terms of  the (0$_1$, $f_{7/2}$)  amplitude  in 
 Fig.~\ref{weight}. The $2^+$ branching ratio is almost completely determined
by the  (2$_1$, $f_{7/2}$) partial width; the second-order contribution from the 
$p_{3/2}$ wave is much smaller (cf. inset in Fig.~\ref{weight}).

\section {Conclusions}\label{SecCon}

This work  contains the first application of 
the triaxial
non-adiabatic weak coupling approach to 
 the description of  proton-emitting nuclei. The resulting
coupled-channel equations take into account the  coupling to  the 
 $K$=2  band representing collective $\gamma$ vibrations. 
 
 The inclusion of  the $\gamma$ band into the weak-coupling formalism increases
the number  of the coupled channel equations significantly. This makes it very 
difficult to  solve accurately 
the multitude of coupled differential equations with Gamow boundary 
condition. In order to overcome this difficulty, 
 we developed a new formalism, dubbed RMHO, which incorporates 
  the variational 
oscillator expansion method
into  the R-matrix theory. Within RMHO, it is possible to  significantly increase
the number of states in the daughter nucleus to guarantee
the convergence of the solution.

As an example, the RMHO 
formalism has been  applied to the g.s. proton emission from
$^{141}$Ho, in which 
there have been   some experimental hints (e.g., 
large signature splitting in the g.s.
rotational band or presence of 
low-lying $\gamma$-vibrational states in the neighboring even-even nuclei)
for triaxiality. 
Our  calculations show that while 
the coupling to $\gamma$ vibrations can in general  influence
decay characteristics (half-life, branching ratios), in the case  of
$^{141}$Ho the resulting trend is opposite to what  has been observed
 experimentally. 
From this point
of view, our results support conclusions drawn in the recent work 
\cite{dav03} based on the adiabatic particle-rotor approach. An important
piece of physics which is still missing in our  non-adiabatic formalism 
is the inclusion of quasi-particle pairing. We are currently working
on incorporating  the Hartree-Fock-Bogoliubov 
couplings  \cite{Bel87} into our model.

\begin{acknowledgments}

Useful discussions with K. Rykaczewski are gratefully acknowledged.
This research was supported by Hungarian OTKA Grant Nos. T37991 and T046791,  
the NATO grant PST CLG.977613 and by
the U.S.\ Department of Energy
under Contract Nos.\ DE-FG02-96ER40963 (University of Tennessee),
 DE-AC05-00OR22725
with UT-Battelle, LLC (Oak Ridge National Laboratory), and 
DE-FG05-87ER40361 (Joint Institute for Heavy
Ion Research).

\end{acknowledgments}

\appendix 

\section{Triaxial  Coulomb potential}
The Coulomb interaction between the proton and the daughter nuclei is
\begin{equation}
V_C({\bf r})=\int \frac{\rho ({\bf r}')}{\vert {\bf r}-{\bf r}'\vert} 
d{\bf r}',
\end{equation}
where the charge density reads 
\begin{equation}
\rho ({\bf r})=\frac{\rho_0}{1+\exp \left[(r-R(\Omega))/a\right]}
\end{equation}
and the nuclear surface  $R(\Omega)$ is given by Eq.~(\ref{surf}).
To the first order in $a_2$ the Coulomb potential is
\begin{equation}
V_C({\bf r})=V^{(1)}_C({\bf r})\vert_{a_2=0}+a_2 V^{(2)}_C({\bf r}).
\end{equation}
The first term, $V^{(1)}_C$,  is the Coulomb potential due 
to an axial symmetric charge density. It is given by a simple expression
derived in, e.g.,  Ref.~\cite{defcou}. The second term, $V^{(2)}_C$,
is of the form
\begin{equation}
V^{(2)}_C({\bf r})=\int \frac{\partial \rho ({\bf r}')}{\partial a_2}
\frac{1}{\vert {\bf r}-{\bf r}'\vert} 
d{\bf r}'.
\end{equation}
Calculating the derivative of the charge density and taking the limit 
of the  sharp  
charge distribution
($a\rightarrow 0$), one obtains  
\begin{equation}
V^{(2)}_C({\bf r})=R_0\rho_0 C(a_0,a_2) \int 
\frac{R^2_a(\Omega')[Y_{2,2}(\Omega')+Y_{2,-2}(\Omega')]}
{\sqrt{ R^2_a(\Omega')+r^2-2R_a(\Omega')r\cos\xi }}d\Omega',
\end{equation}
where
\begin{equation}
\cos\xi=\cos \theta \cos\theta'+\sin\theta\sin\theta'\cos(\phi-\phi'). 
\end{equation}
and $R_a(\Omega')=R(\Omega')\vert_{a_2=0}$.
This can be reduced to 
\begin{eqnarray}
&V^{(2)}_C({\bf r})=R_0\rho_0C(a_0,a_2)\sqrt{\frac{15}{8\pi}}\nonumber\\
& \int_{-1}^1dtR^2_a(t)(1-t^2)
\int_0^{2\pi}d\phi'\frac{2\cos^2\phi'-1}
{\sqrt{R^2_a(t)+r^2-2R_a(t)r\cos\xi}},
\end{eqnarray} 
where $t=\cos\theta'$. The integral over $\phi'$ can be calculated 
analytically, 
and the  final result can be expressed in terms of a simple 
one-dimensional integral:
\begin{eqnarray}\label{intt}
&V^{(2)}_C({\bf r})=R_0\rho_0C(a_0,a_2)\sqrt{\frac{15}{8\pi}}
\cos({2\phi})\nonumber\\
& \int_{-1}^1dtR^2_a(t)(1-t^2)\frac{4}{3b^2\sqrt{a+b}}
\left \{ (4a^2-b^2)K(\kappa)-4a(a+b)E(\kappa)\right \},
\end{eqnarray} 
where
\begin{equation}
a=R_a^2(t)+r^2-2R_a(t)r\cos\theta\cos\theta',
\end{equation}
\begin{equation}
b=2R_a(t)r\sin\theta\sin\theta',
\end{equation}
and
\begin{equation}
\kappa=\sqrt{\frac{2b}
{a+b}}.
\end{equation}
In Eq.~(\ref{intt}),  $K(\kappa)$ and $E(\kappa)$ are 
the complete elliptic integral of the first and second kind, respectively.

\end{document}